\documentclass{article}
\usepackage{spconf,amsmath,graphicx,booktabs, amssymb}

\usepackage{enumitem}
\setlist{nosep, leftmargin=14pt}
\usepackage{mwe} 
\usepackage{amssymb}  
\usepackage{graphicx} 
\usepackage{booktabs} 
\usepackage{array}    
\usepackage{tabularx}  
\usepackage[colorlinks, urlcolor=black, linkcolor=blue, citecolor=blue]{hyperref}

\title{Salient Region Matching for Fully Automated MR-TRUS registration}
\name{Zetian Feng, Dong Ni, Yi Wang$^{*}$
\thanks{$*$ Corresponding author}
\thanks{This work was supported in part by the National Natural Science Foundation of China under Grants 62471306 and 62071305,
in part by the Shenzhen Medical Research Fund under Grant D2402010,
in part by the Guangdong-Hong Kong Joint Funding for Technology and Innovation under Grant 2023A0505010021,
in part by the Guangdong Basic and Applied Basic Research Foundation under Grant 2022A1515011241,
and in part by the Shenzhen Major Scientific and Technological Project (KJZD20230923114615031).}
}
\address{
	School of Biomedical Engineering,
	Shenzhen University, Shenzhen, China,\\
	and the Medical UltraSound Image Computing (MUSIC) Lab, Shenzhen, China,\\
	and the Smart Medical Imaging, Learning and Engineering (SMILE) Lab, Shenzhen, China\\}
\begin{document}
%
\maketitle
\begin{abstract}
Prostate cancer is a leading cause of cancer-related mortality in men.
The registration of magnetic resonance (MR) and transrectal ultrasound (TRUS) can provide guidance for the targeted biopsy of prostate cancer.
In this study, we propose a salient region matching framework for fully automated MR-TRUS registration.
The framework consists of prostate segmentation, rigid alignment and deformable registration.
Prostate segmentation is performed using two segmentation networks on MR and TRUS respectively, and the predicted salient regions are used for the rigid alignment.
The rigidly-aligned MR and TRUS images serve as initialization for the deformable registration.
The deformable registration network has a dual-stream encoder with cross-modal spatial attention modules to facilitate multi-modality feature learning, and a salient region matching loss to consider both structure and intensity similarity within the prostate region.
Experiments on a public MR-TRUS dataset demonstrate that our method achieves satisfactory registration results, outperforming several cutting-edge methods.
The code is publicly available at~\textit{https://github.com/mock1ngbrd/salient-region-matching}.
\end{abstract}
\section{Introduction}
\label{sec:intro}
Prostate cancer is one of the most prevalent cancers among men worldwide~\cite{sung2021global}.
Transrectal ultrasound (TRUS) is favored for its real-time imaging and cost-effectiveness for the guidance of prostate interventions, 
though its limited tissue contrast and resolution hinder precise lesion identification.
Conversely, magnetic resonance (MR) imaging offers high-quality soft tissue contrast, making it ideal for indicating lesions but inefficient for real-time guidance.
MR-TRUS registration leverages the complementary advantages of both modalities, enhancing diagnostic accuracy and improving targeted interventions.

However, a major challenge in multi-modality image registration, particularly between MR and TRUS, is the lack of robust similarity metrics.
To address this, various deep learning methods have been proposed, including image-based methods, point set methods, and label-driven methods~\cite{wang2024review}.
Image-based methods~\cite{yan2018adversarial, haskins2019learning, song2022cross} mainly optimize the loss function that calculates the similarity between MR-TRUS images.
Yan~\textit{et al}.~\cite{yan2018adversarial} proposed an adversarial framework by training two convolutional neural networks (CNNs) simultaneously, one being a generator for image registration and the other being a discriminator as a similarity metric.
Haskins~\textit{et al}.~\cite{haskins2019learning} used a CNN to learn the similarity metric through the registered MR-TRUS pairs.
Song~\textit{et al}.~\cite{song2022cross} introduced a cross-modal attention module to establish feature correspondence.
However, most image-based methods focused on rigid registration or needed ground truth transformation for supervision.
Point set methods~\cite{baum2020multimodality, baum2021real, fu2021biomechanically, min2023non} reframe image registration as a point cloud matching problem, bypassing intensity-based similarity measures by aligning prostate surfaces.
Some methods~\cite{baum2020multimodality, baum2021real} estimated deformation fields only relying on the point cloud generated from prostate surface, resulting in inadequate inner deformation.
To deal with this issue, biomechanical constraints have been added~\cite{fu2021biomechanically, min2023non}, but computational cost increased meanwhile.
Label-driven methods~\cite{hu2018label, hu2018weakly, zeng2020label, chen2021mr} leverage annotated corresponding structures to evaluate the shape similarity.
These structures could be prostate gland, vessel, point landmarks, etc.
Hu~\textit{et al}.~\cite{hu2018label} proposed a weakly-supervised method, learning voxel correspondence based on higher level label correspondence.
Later, Hu~\textit{et al}.~\cite{hu2018weakly} extended their work by designing a multi-scale Dice loss and an improved network architecture.
Zeng~\textit{et al}.~\cite{zeng2020label} trained two separate CNNs for MR and TRUS segmentation, and employed the predicted prostate masks to align MR-TRUS images.
Similarly, Chen~\textit{et al}.~\cite{chen2021mr} used segmentation-based learning for MR-TRUS registration.
However, \cite{zeng2020label, chen2021mr} only focused on the gland similarity and failed to account for internal structure alignment.
Annotating corresponding internal structures for weakly-supervised learning~\cite{hu2018label, hu2018weakly} requires medical expertise.

By reviewing and rethinking various MR-TRUS registration methods,
we consider a successful solution should address following issues:
(1) the method has to pay more attention to the foreground (i.e., prostate) registration, since MR and TRUS are with different field of view where pelvic MR scans normally contain much more anatomical structures.
(2) A reliable and accurate registration shall consider both structure and intensity similarity, which simultaneously constrains the shape and meanwhile providing dense matching.

In this study, we propose a salient region matching framework for fully automated MR-TRUS registration.
The proposed framework considers both structure and intensity similarity within the prostate region.
Experiments on a public dataset demonstrate the proposed method outperforms cutting-edge methods for MR-TRUS registration.
Our main contributions are summarized as follows:
\begin{itemize}
\item[$\bullet$]
We develop a salient region matching framework, which consists of prostate region-of-interest (ROI) segmentation, rigid alignment, and deformable registration, providing accurate and automated MR-TRUS registration.

\item[$\bullet$]
To facilitate multi-modality feature learning, we design a dual-stream registration encoder to capture uncoupled features of each modality, and cross-modal spatial attention modules to refine features with spatial information from the other modality.

\item[$\bullet$]
To constrain the effective foreground registration, we propose a salient region matching loss, which evaluates the structure and intensity similarity within the prostate ROI.
\end{itemize}

\section{Method}
\label{sec:method}

\begin{figure}[t]
	\centering
	\includegraphics[width=\linewidth]{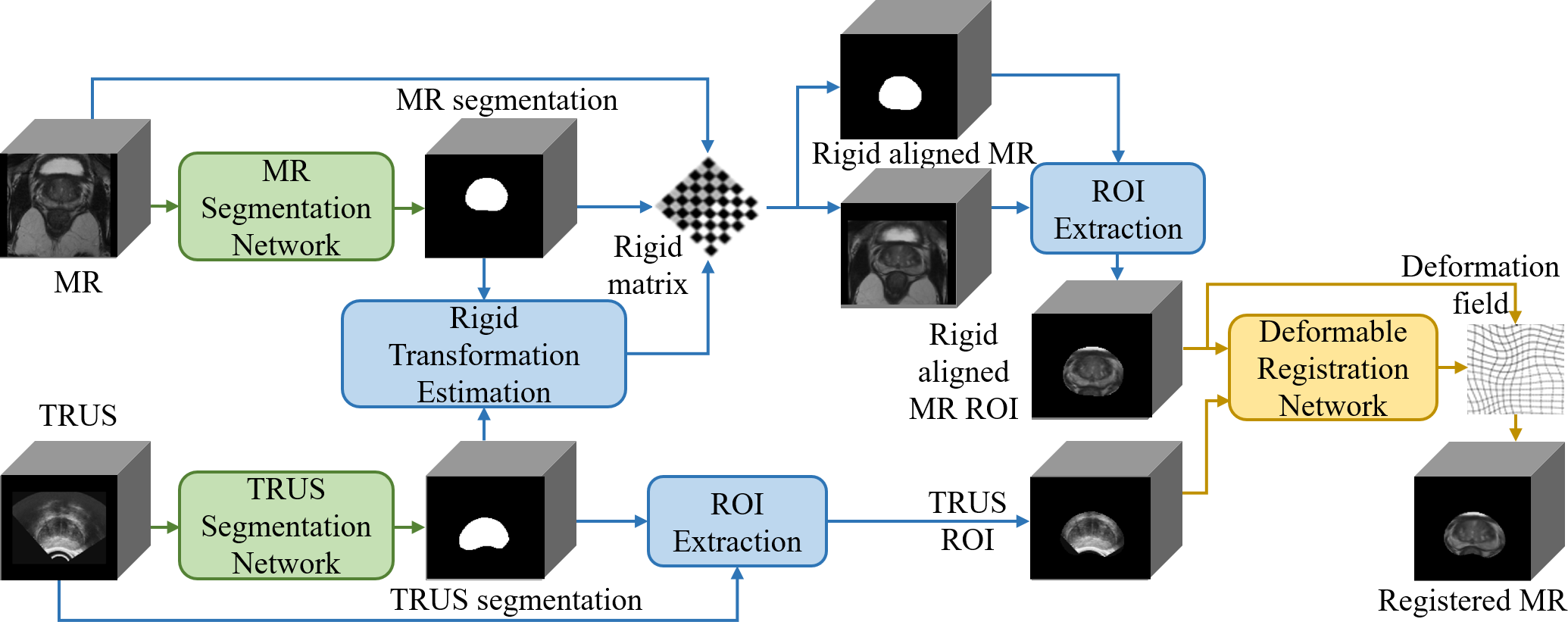}
	\caption{Illustration of our MR-TRUS registration framework.}
	\label{fig: framework}
\end{figure}

\subsection{Framework Overview}
\label{ssec:method1}
Fig.~\ref{fig: framework} shows the proposed MR-TRUS registration framework, which consists three main components including prostate segmentation, rigid alignment, and deformable registration.
Below, we discuss the specific workflow of each component.

1)~\textbf{ROI Segmentation.}
The framework begins with prostate region segmentation, where two V-Nets~\cite{milletari2016v} are used to segment the prostate in MR and TRUS images, respectively.
This step efficiently localizes the prostate region, facilitating subsequent rigid and deformable registration.

2)~\textbf{Rigid Alignment.}
Predicted prostate regions (i.e., binary masks) are utilized to unify the multi-modality images into the same global coordinate.
Specifically, rigid alignment is performed on prostate masks using ANTS~\cite{ants2009} to obtain the rigid transformation matrix.
The rigid matrix is employed to deform the MR image and the predicted MR mask.
Then the prostate ROIs are extracted from TRUS and rigid aligned MR images based on segmentation results.
This process removes redundant background information and provides globally-aligned salient regions for the deformable registration.

3)~\textbf{Deformable Registration.}
The registration network has a encoder-decoder structure and regresses the volumetric deformation field for the final registration.
More details are discussed in Section~\ref{ssec:method2}.

\subsection{Deformable Registration Network}
\label{ssec:method2}
\subsubsection{Network Architecture}

\begin{figure}[t]
	\centering
	\includegraphics[width=\linewidth]{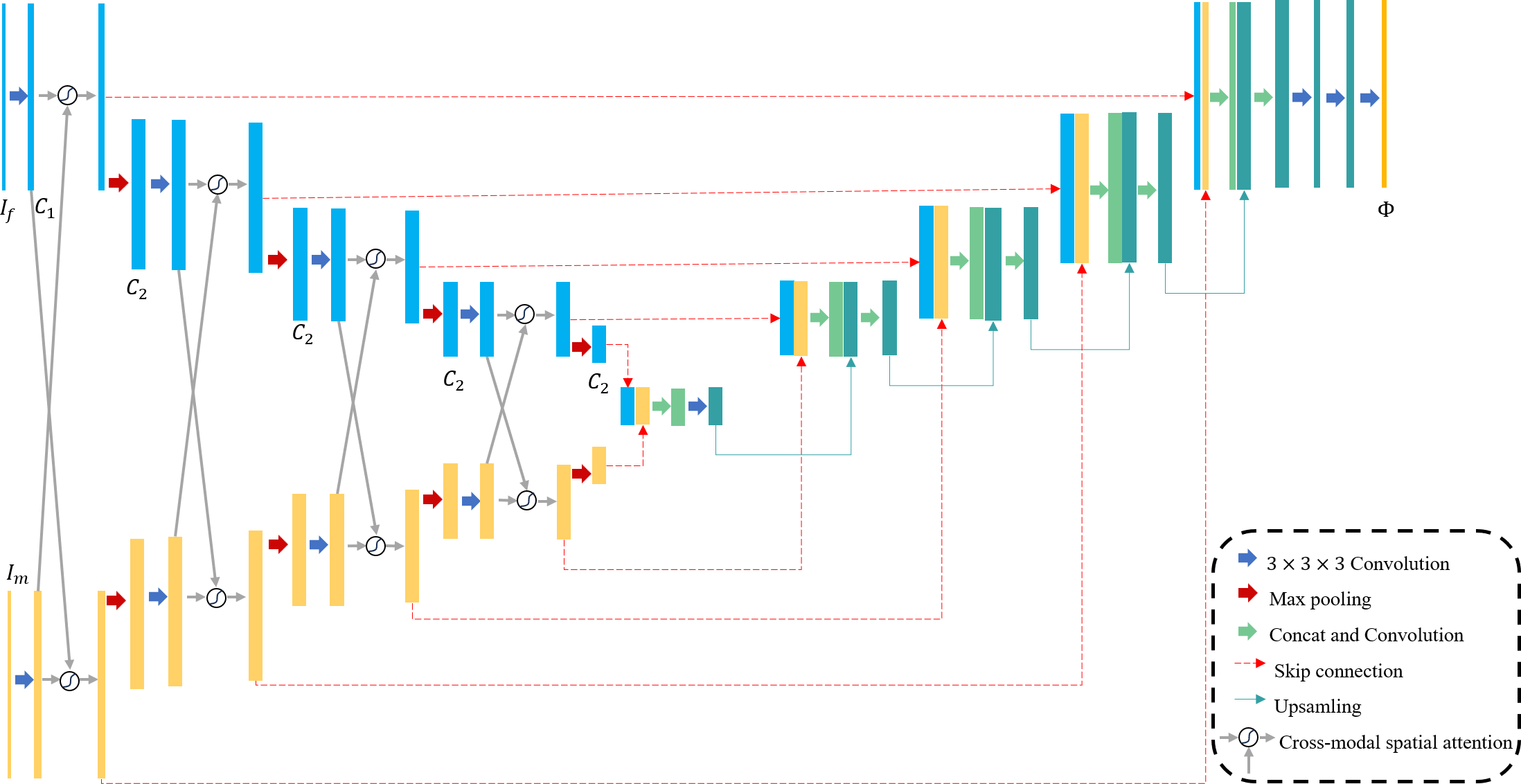}
	\caption{The proposed deformable registration network.}
	\label{fig: network architecture}
\end{figure}

The architecture of our deformable registration network consists of three main components (see Fig.~\ref{fig: network architecture}):
a dual-stream encoder, the cross-modal spatial attention (CMSA) module, and a decoder.
The encoder employs a dual-stream design without parameter sharing, effectively capturing modality-specific features from each modality ($I_f$ and $I_m$ denote the TRUS ROI and rigid aligned MR ROI, respectively). 
Each stream has four down-sampling levels, with each level including a convolutional block, a CMSA module and a max pooling layer.
The convolutional block contains a \(3\times3\times3\) convolutional layer and a LeakyReLU activation layer.
The encoder starts with \({C}_1= 16\) channels and doubles them at the second level, while halving the feature map size at each level.
In the decoder, this process is reversed with upsampling blocks.
Each upsampling block contains a concatenation and convolutional block, along with a upsampling layer.
Cross-modal fusion skip connections merge feature maps from both streams via concatenation and convolutional blocks, passing the fused features to the corresponding upsampling block at each level.
These skip connections preserve high-resolution spatial information, improving spatial correspondence decoding.

\begin{figure}[t]
\centering
\includegraphics[width=\linewidth]{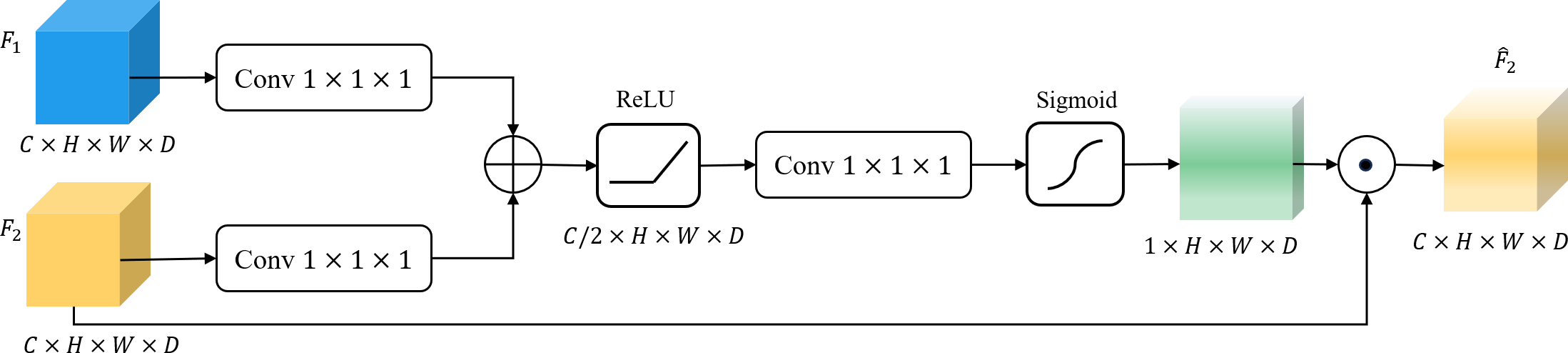}
\caption{The cross-modal spatial attention (CMSA) module.}
\label{fig: CMSA}
\end{figure}

Inspired by~\cite{oktay2018attention}, the CMSA module, serving as a bridge between the two encoding streams at each level, is designed to enhance spatial features by leveraging information from the other modality.
As shown in Fig.~\ref{fig: CMSA}, the module aggregates feature maps from both modalities to generate a spatial attention map for the current stream.
This attention map is then multiplied element-wise with the feature maps, emphasizing relevant regions while suppressing irrelevant ones.
This mechanism allows the network to learn multi-modality feature representation.
The cross-modal spatial attention is formulated as follows:
\begin{align}
\label{eq: cmsa}
att =  \sigma_2(W_3^T\sigma_1(&W_1^T F_1 + W_2^T F_2 + b_{1,2})+b_3),\\
&\hat{F_2} = att \odot F_2,
\end{align}
where \(F_1\) and \(F_2\) are features from different modalities, $att$ is spatial attention map, \(\hat{F_2}\) is the final features refined by element-wise multiplying \(F_2\) with $att$.
\( W_1 \in \mathbb{R}^{C \times \frac{C}{2}} \),  \( W_2 \in \mathbb{R}^{C \times \frac{C}{2}} \), and \( W_3 \in \mathbb{R}^{\frac{C}{2} \times 1 } \) are linear transform matrices, \(b_{1,2}\) and \(b_3\) are bias terms.
The linear transformations are implemented as channel-wise \(1\times 1\times1\) convolutions.
\(\sigma_1\) and \(\sigma_2\) are ReLU and Sigmoid activation function, respectively.

\subsubsection{Salient Region Matching Loss}
\label{ssec: ISHL}

We design salient region matching loss (SRML) to consider both intensity and structure similarity within the prostate region.
The mutual information (MI)~\cite{563664} computed within the prostate ROI constrains the network to focus on the internal structures, reducing distractions from irrelevant anatomical information outside the prostate.
The MI loss is formulated as:
\begin{equation}
\label{eq: mi}
\mathcal{L}_{\text{ROI-MI}} = - \text{MI}(I_{\text{MR}}^{\text{r}}, I_{\text{TRUS}}^{\text{r}}),
\end{equation}
where \(I_{\text{MR}}^{\text{r}}\) and \(I_{\text{TRUS}}^{\text{r}}\) represent the prostate ROIs in MR and TRUS images, respectively. 
The \(\text{MI}(X, Y)\) measures the shared information between two intensity distributions \(X\) and \(Y\)~\cite{563664}, capturing both the correlations across modalities.

Since MI primarily captures intensity distribution similarities between images, it lacks the ability to fully account for anatomical structure alignment.
Therefore, we introduce a weighted multi-class Dice loss to incorporate structural information.
By leveraging structural information from annotations of various anatomical structures, the Dice loss enforces the network to focus on aligning these structures across modalities.
The multi-class Dice loss is defined as:
\begin{equation}
\label{eq: dice}
\mathcal{L}_{\text{ROI-Dice}} = 1 - \frac{1}{K} \sum_{k=1}^{K} w_k \cdot \frac{2 |P_k \cap G_k|}{|P_k| + |G_k|},
\end{equation}
where \(K\) is the number of structure classes (e.g., prostate gland, lesions, etc),
\(P_k\) and \(G_k\) represent the corresponding structure masks in MR and TRUS for class \(k\),
and \(w_k\) is the weight assigned to class \(k\), reflecting the importance of each class in guiding the registration process.

The $L_2$-norm of the gradient of deformation field is employed to regularize unrealistic deformations.
Given a deformation field $\phi$ and the voxel \(v\) in the whole volume \(\Omega\), the regularization loss is defined as:
\begin{equation}
\label{eq: reg}
\mathcal{L}_{\mathrm{reg}}(\phi)=\sum_{v \in \Omega}\|\nabla \phi(v)\|^2.
\end{equation}

The overall training loss function is:
\begin{equation}
\mathcal{L}_{\text{train}} = \mathcal{L}_{\text{ROI-MI}} + \mathcal{L}_{\text{ROI-Dice}} + \lambda \mathcal{L}_{\mathrm{reg}}(\phi),
\end{equation}
where \(\lambda\) is weighting factor for regularization term.

\section{Experiments}
\label{exp}
\subsection{Dataset}
\label{d}
Experiments were conducted on a public dataset, $\mu$-RegPro \cite{baum2023mr}.
MR images were acquired using 1.5T/3T scanners at UCLH, with all volumes standardized to \(120\times128\times128\) voxels.
TRUS volumes were reconstructed from 57-112 frames obtained by rotating a bi-plane probe, yielding volumes of \(81\times118\times 88\) voxels.
All images were resampled to \(0.8mm^3\) isotropic voxel size,
and center-padded to \(128\times128\times128\) voxels.
The dataset was divided into training and testing sets, with 58 and 15 MR-TRUS pairs, respectively.

Each pair of MR and TRUS volumes included prostate gland segmentation and corresponding anatomical landmark annotations.
The landmarks included lesions, zonal structures, cysts, calcifications, and specific patient landmarks such as vas deferens and seminal vesicles, etc.

\subsection{Implementation Details}
\label{ssec: implementation details}
For segmentation network training, the batch size was set to 5, using Adam optimizer with an initial learning rate of $10^{-3}$.
For registration network training, the batch size was set to 1, using Adam optimizer with an initial learning rate of $10^{-4}$.
We set \(w_k\) to 0.1 for the prostate gland and 0.3 for other landmarks, and \(\lambda\) to \(0.4\).
All networks were implemented in Pytorch on a single NVIDIA GeForce RTX 2080 Ti GPU.
Rigid alignment utilized the ANTsPy library, employing mean square error as the similarity metric with default settings for other parameters.

\subsection{Comparison Methods and Ablation Study}
\label{ssec: comparison and abla}
We compared our method with two state-of-the-art label-driven methods:
a weakly-supervised network using prostate mask and all other landmarks as supervision signals~\cite{hu2018weakly},
and a segmentation-based method using prostate segmentation results to conduct registration~\cite{chen2021mr}.
Note that~\cite{hu2018weakly} did not contain rigid alignment, while~\cite{chen2021mr} included this operation.
We also compared~\cite{hu2018weakly} with rigid initialization.
Considering landmark annotations are labor-intensive,
we further compared two variants of our method that used (1) only prostate mask, (2) prostate mask and all other landmarks in Eq~(\ref{eq: dice}).

Ablation studies on each component were also conducted.
First, we evaluated our method with and without rigid alignment.
Second, we examined the impact of the CMSA.
Lastly, we performed ablation analysis on the SRML, where we only used Dice loss for optimization instead, similar to~\cite{hu2018label}.

\subsection{Evaluation Metrics}
\label{ssec: em}
Dice similarity coefficient (DSC) and target registration error (TRE) were used to evaluate the registration accuracy.
TRE is defined as the root mean square of the distance error between centroids of landmark pairs,
while DSC measures the overlap between the prostate glands in TRUS and registered MR.
Larger DSC and smaller TRE indicate better registration.


\begin{table}[t]
	\centering
	\caption{Quantitative results of different methods.}
	\label{tab: rw}
	\footnotesize
	\begin{tabular}{ccc}
			\toprule
			Methods & DSC (\%) & TRE (mm) \\
			\hline
			Initial & 59.4$\pm$17.6 & 11.45$\pm$3.81 \\
			\hline
			Rigid & 80.4$\pm$5.1 & 5.49$\pm$2.13 \\
			\cite{hu2018weakly} & 70.9$\pm$16.2 & 8.63$\pm$4.05 \\
			Rigid+\cite{hu2018weakly} & 81.9$\pm$5.5 & 4.99$\pm$2.26 \\
			\cite{chen2021mr} & 83.4$\pm$5.5 & 5.29$\pm$2.25 \\
			\hline
			Ours (1) & 85.0$\pm$3.7 & 5.18$\pm$2.09 \\
			Ours (2) & \textbf{85.9$\pm$3.5} & \textbf{4.65$\pm$1.76} \\
			\bottomrule
	\end{tabular}
\end{table}

\begin{table}[t]
	\centering
	\caption{Quantitative results of the ablation study.}
	\label{tab: abla}
	\footnotesize
	\begin{tabular}{cccccc}
			\toprule
			Rigid & CMSA & SRML & DSC (\%) & TRE (mm) \\
			\hline
			&  \(\checkmark\) & \(\checkmark\) & 81.4$\pm$8.2 & 7.09$\pm$3.01 \\
			\(\checkmark\) & & \(\checkmark\) & 80.3$\pm$5.0 & 5.48$\pm$2.10 \\
			\(\checkmark\) & \(\checkmark\) & & 80.4$\pm$4.6 & 5.00$\pm$1.98 \\
			\(\checkmark\) & \(\checkmark\) & \(\checkmark\) & \textbf{85.9$\pm$3.5} & \textbf{4.65$\pm$1.76} \\
			\hline
	\end{tabular}
\end{table}

\begin{figure}[t]
	\centering
	\centerline{\includegraphics[width=0.89\linewidth]{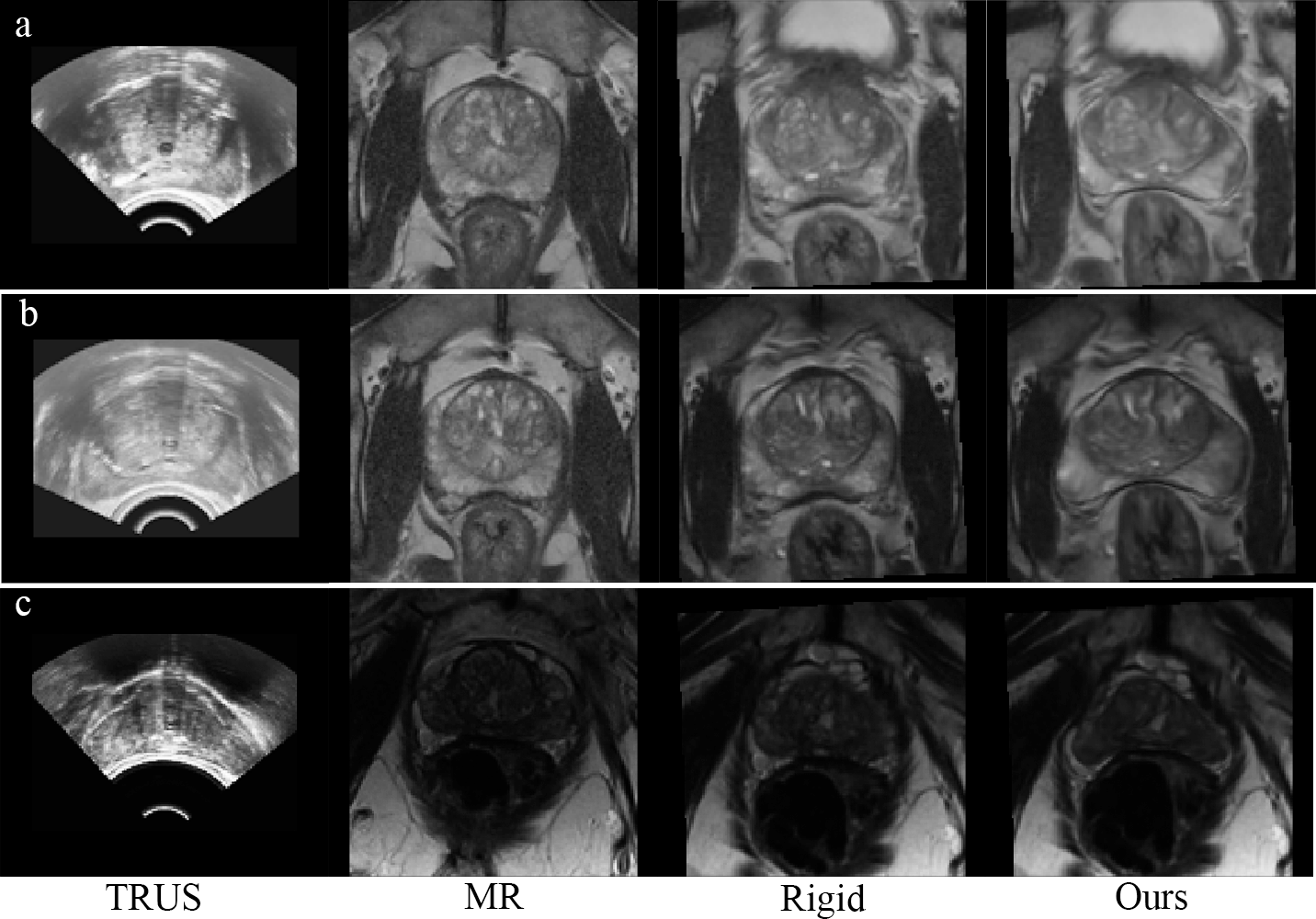}}
	\caption{Visualization of the MR-TRUS registration results.}
	\label{fig: results}
\end{figure}

\subsection{Experimental Results}
\label{ssec: er}
The performance of prostate segmentation was first verified. 
The networks achieved satisfactory performance in both modalities, DSC of 88.6\% for MR and 90.4\% for TRUS.

Table~\ref{tab: rw} reports the comparison results of different methods.
The ``Rigid'' method refers to the rigid alignment in our approach.
It can be observed that our registration method using all landmarks outperformed other methods in terms of DSC and TRE.
Specifically, it surpassed~\cite{hu2018weakly} even with the same rigid alignment as initialization, highlighting the superiority of our registration architecture and salient region matching mechanism.
Compared to~\cite{chen2021mr}, which excelled in surface registration, our method also attained better alignment on prostate surface and internal structures.
Additionally, our method using only prostate mask also outcompeted~\cite{hu2018weakly, chen2021mr}.
It had larger TRE compared to ``Rigid+\cite{hu2018weakly}'' due to the lack of internal landmarks' guidance.
Fig.~\ref{fig: results} shows some registration results.
Our method effectively handled large and complicated deformations between MR and TRUS images.

Table~\ref{tab: abla} lists the qualitative results of the ablation study.
Rigid alignment largely improved both DSC and TRE, underscoring the importance of effective initialization before deformable registration.
Removing the CMSA module resulted in DSC and TRE values equivalent to those of rigid alignment.
This highlights the necessity of the CMSA module for enabling the dual-stream encoder to better capture multi-modality feature representation.
Without SRML,
the network struggled to register the prostate surface without the consideration of the hybrid structure-intensity similarity.

\section{Conclusion}
\label{sec: conclusion}
MR-TRUS registration has great clinical significance yet remains challenges.
We propose a salient region matching framework for fully automated MR-TRUS registration.
The framework first localizes prostate ROIs then conducts rigid initialization.
The deformable network employs CMSA to learn multi-modality feature representation, and leverages SRML to constrain the hybrid structure-intensity similarity.
Experiments have proven the favorable performance of our proposed MR-TRUS registration framework.
Considering the used dataset is with limited images, we attempt to evaluate our method on larger and independent datasets to validate its generalizability.
In addition, the separate segmentation process may limit the achievable registration performance.
Future work may focus on joint segmentation and registration, which simultaneously optimizes these two tasks.\\

\noindent\textbf{Compliance with Ethical Standards}\\
This research study was conducted retrospectively using medical imaging data made available in open access by~\cite{baum2023mr}.
Ethical approval was not required as confirmed by the license attached with the open access data.

\hspace*{\fill}\\

\noindent\textbf{Conflicts of Interest}\\
The authors have no conflicts to disclose.


\bibliographystyle{IEEEbib}
\bibliography{refs}

\end{document}